\documentstyle[aps,epsbox]{revtex}
\begin{document}
\tightenlines
\title{Noise reduction in signal processing using binary couplings}
\author{Jun-ichi Inoue 
and Domenico M. Carlucci }
\address{Department of Physics, 
Tokyo Institute of Technology, 
Oh-okayama 
Meguro-ku
Tokyo 152, 
Japan 
}
\maketitle
\begin{abstract}
We report a simple extension of a model for noise reduction 
in signal processing, already introduced by Mourik {\it et al.}, 
in the presence of  binary coupling vectors, which turns to be more useful 
for practical and engineering implementations. We also compute 
annealed approximation which gives an upper bound of the correct 
critical number of noise-sources. Finally, we also find that the full RSB 
Parisi solution just above the AT line is an universal function for any 
symmetric distribution of the coupling vectors. 
\end{abstract}
\vspace{1cm}

\section{The model}

In a recent paper\cite{Mourik}, Mourik {\it et al.} investigated the 
statistical properties for the problem of noise reduction (NR) in signal 
processing. For this model one considers $N$-detectors receiving 
a signal mixed with noise from $p$ sources. The total input on a source 
from the detector $j$ reads 
\begin{equation}
x_{j}=a_{j}S+\sum_{\mu}{\xi}_{j}^{\mu}
n_{\mu}
\label{detect}
\end{equation}
where $S$ and $n_{\mu}$ are respectively the amplitudes of signal and 
of the noise from $\mu$-th source, ${\xi}_{j}$'s are 
random gaussianly distributed variables, and $a_j$'s fixed numbers. 
The goal is to find a linear combination of the inputs in order to reduce 
the noise and at the same time have a clear signal. 
By using a $N$-dimensional weight vector as  a ``filter'',  
the sum of observable inputs $x_{j}$ weighted with such a vector, reads    
\begin{equation}
h\,{\equiv}\,\frac{1}{\sqrt{N}}
\sum_{j}J_{j}x_{j}=\frac{S}{\sqrt{N}}
\sum_{j}J_{j}a_{j}+\sum_{\mu}\sum_{j}
\frac{J_{j}{\xi}_{j}^{\mu}}
{\sqrt{N}}.
\label{filter}
\end{equation}
Now, in order to reduce the noise from the local field 
$h^{\mu}=\sum_{j}J_{j} \xi_{j}^{\mu}/\sqrt{N}$ 
and have the signal part $S\sum_{j}J_{j}a_{j}/\sqrt{N}$ 
constant ${\sim}{\cal O}(1)$, we use the strategy which 
directly minimize the quantity $|h^{\mu}|$. 
If the number of the noise sources $p$ is small, 
it is straightforward to find the optimal solution which minimizes the 
``cost" $|h^{\mu}|$. 
However, as the number of the noise sources 
increases, in general, it becomes hard to 
solve such optimization problem. 
In this case, it is natural to wonder  
how many configurations $\mbox{\boldmath $J$}={\pm}1$ 
do exist when the phase space is restricted to $|h^{\mu}|<k$, where $k$ is the 
noise tolerance. For each configuration of the noise, the number of 
the solutions $\mbox{\boldmath $J$}$ which satisfy the previous condition 
can be expressed as  
\begin{equation}
{\cal N}(\mbox{\boldmath $J$})=
{{\rm Tr}_{\{\mbox{\boldmath $J$}\}}}
{\prod}_{\mu=1}^{p}{\Theta}(k^{2}-(h^{\mu})^{2})
\label{number}
\end{equation}
and the ``entropy" of the solution space is given by the its logarithm 
${\cal S}(\mbox{\boldmath $J$})={\log}{\cal N}(\mbox{\boldmath $J$})$. 

In order to investigate the typical properties of the entropy, 
one usually averages the above quantity 
with respect to the gaussian noise source $\mbox{\boldmath ${\xi}$}$, 
namely
\begin{equation}
<{\cal S}(\mbox{\boldmath $J$})>_{\mbox{\boldmath $\xi$}}
=<{\log}{\cal N}(\mbox{\boldmath $J$})>_{\mbox{\boldmath $\xi$}}.
\label{average}
\end{equation}

The previous quenched average can be performed by means of the usual replica  
method\cite{Parisi}, thus giving

\begin{equation}
<{\cal N}^{n}(\mbox{\boldmath $J$})>_{\mbox{\boldmath $\xi$}}=
<{\rm Tr}_{\{\mbox{\boldmath $J$}\}}
{\prod_{\mu=1}^{p}}{\prod_{a=1}^{n}}
{\Theta}(k^{2}-(h_{a}^{\mu})^{2})
>_{\mbox{\boldmath $\xi$}}.
\end{equation}

Therefore, after performing the average with respect to the gaussian 
randomness and introducing auxiliary variables of integrations, 
the free energy simply reads 

\begin{eqnarray}
\mbox{} & \mbox{} & \exp[f]  \sim  \int \prod_{\alpha<\beta}dq_{\alpha\beta}d\tilde{q}_{\alpha\beta} \nonumber \\
\mbox{} & \times & \exp\left[N\,\left( 
	           \alpha 
		   {\cal G}_0(\{q\}) 
	           \,+\,
                   {\cal G}_1(\{\tilde{q}\})
                   \,+\, 
                   \frac{1}{2} 
                   \sum_{\alpha\neq\beta} 
                   q_{\alpha\beta} \tilde{q}_{\alpha\beta}
               \right)
       \right]
\label{replica_free_energy}
\end{eqnarray}

where 

\begin{eqnarray} 
      & \mbox{} & \exp\left[ {\cal G}_0(\{q\}) \right] \nonumber \\ 
     \mbox{} & = & 
     \int_{-\kappa}^{\kappa} \left[d\lambda\right]
     \int_{-\infty}^{\infty} \left[d h\right]
     \exp\left[ 
               -\frac{1}{2}\sum_{\alpha} h_{\alpha}^2
               -
               \frac{1}{2}\sum_{\alpha\neq\beta} 
               h_{\alpha}q_{\alpha\beta} h_{\beta}
         \right]
\end{eqnarray} 

and

\begin{equation} 
     \exp\left[ {\cal G}_1(\{\tilde{q}\}) \right] 
     \,=\,
     {\rm Tr}_{\{\mbox{\boldmath $J$}\}}\,
     \exp\left[ 
               \frac{1}{2}\sum_{\alpha\neq\beta} 
               J_{\alpha}\tilde{q}_{\alpha\beta} J_{\beta}
         \right].
\end{equation} 

As usual, the previous quantities should be evaluated in the limit 
$N\to \infty$, by making an ansatz for the matrices $q_{\alpha\beta}$ 
and $\tilde{q_{\alpha\beta}}$. In the next section, we consider 
the replica symmetric solution.

\section{RS solution}

Within the replica symmetric ansatz 
\begin{equation}
q_{\alpha\beta}{\equiv}\frac{1}{N}\sum_{j}J_{j}^{\alpha}J_{j}^{\beta} = q 
\hspace{1cm}
\tilde{q}_{\alpha\beta} =  \tilde{q} \,\,\,\,\,\,\,\,
\forall \alpha,\beta
\end{equation}
it is straightforward to check that 
the entropy per system size $N$ reads, in the limit $n\to 0$

\begin{eqnarray}
\mbox{} & \mbox{} & \frac{{\log}<{\cal N(\mbox{\boldmath $J$})}>_{\mbox{\boldmath $\xi$}}}
{Nn}{\equiv}{\cal S}(q,\tilde{q}) \nonumber \\ 
\mbox{} & = &   
{\rm extr}_{q,\tilde{q}}{\Bigr [}{\alpha}\int{Dy}
{\log}
\left[
H\left(
\frac{-k+\sqrt{q}y}
{\sqrt{1-q}}
\right)
-
H\left(
\frac{k+\sqrt{q}y}
{\sqrt{1-q}}
\right)
\right] \nonumber \\
\mbox{} & - & 
\frac{\tilde{q}}{2}(1-q) 
+ \int{Dt}{\log}2{\cosh}(\sqrt{\tilde{q}}t){\Bigr ]}.
\end{eqnarray}
The stationary points of the entropy with respect to $q$ and $\tilde{q}$

\begin{equation}
\frac{\partial {\cal S}(q,\tilde{q})}
{\partial q}=
\frac{\partial {\cal S}(q,\tilde{q})}
{\partial \tilde{q}}=0.
\label{saddle}
\end{equation}

give the optimal number of the noise sources  
as a function of the noise  tolerance $k$.
Moreover, in order to obtain the critical number of the noise source 
${\alpha}_{\rm c}$, beyond which the solution space shrinks down to zero, 
we have to estimate the expression (\ref{saddle}) in the limit 
$q{\rightarrow}1$, {\it viz.}
\begin{equation}
{\alpha}_{c}^{(0)}(k)=\frac{2}{\pi}\left[
2(1+k^{2})H(k)-\sqrt{\frac{2}{\pi}}k\,{\exp}(-\frac{k^{2}}{2})
\right]^{-1}
\end{equation}
We thus get the same result as \cite{Mourik} except for a factor 
$2/\pi$, just as in the case of the random pattern storage 
problems \cite{Gardner,Krauth}.
Therefore, one obtains 
${\alpha}_{\rm c}^{(0)}(k)\,{\simeq}\,\sqrt{2/\pi}k^{3}
{\rm e}^{k^{2}/2}/2$
(large $k$ limit) and 
${\alpha}_{c}^{(0)}(k)\,{\sim}\, 2\left(1+4k/\sqrt{2\pi}
\right)/\pi$
(small $k$ limit).

At fixed $k$, it is important to compare the value of the critical 
capacity with the AT line, beyond which the saddle point solution 
turns to be unstable against transverse fluctuations.
Because of the inversion symmetry of the constraints, 
the saddle point equations (\ref{saddle}) are satisfied with  
the solution $q=0$, independent of $\alpha$ and $k$. 
The stability of such solution is determined by studying 
the eigenvalues of the Hessian around  the saddle point solution.  
Therefore, the AT line is found to be
\begin{equation}
{\alpha}_{\rm AT}(k)=
\frac{\pi}{2}
\frac{[1-2H(k)]^{2}}
{k^{2}{\rm e}^{-\frac{k^{2}}{2}}
}
\end{equation}
as in the continuous case. 
So, qualitatively we get the same results as \cite{Mourik} where 
a discontinuous phase transition seems to appear for large $k$.

In order to give an upper bound for $\alpha_{\rm c}(k)$, we also 
investigate the optimal number of the noise source using annealed 
approximation\cite{Krauth}. So we replace the quenched average by the following 
annealed average
\begin{equation}
<{\cal S}(\mbox{\boldmath $J$})>_{\mbox{\boldmath $\xi$}}
={\log}<{\cal N}(\mbox{\boldmath $J$})>_{\mbox{\boldmath $\xi$}}.
\label{annealed}
\end{equation}
Then we introduce the ``magnetization'' 
${\cal M}=\sum_{j}J_{j}/\sqrt{N}$ and 
rewrite the number of the configurations ${\cal N}$ as 
\begin{equation}
{\cal N}(\mbox{\boldmath $J$})={\rm Tr}_{\{\mbox{\boldmath $J$}\}}
{\prod_{\mu=1}^{p}}
{\Theta}(k^{2}-(h^{\mu})^{2}){\delta}
(\sqrt{N}{\cal M}-\sum_{j}J_{j}).
\end{equation}
It should be noticed that in the quenched calculation, 
the contribution of the magnetization ${\cal M}$ in the entropy 
becomes ${\cal O}(\sqrt{N})$  and vanishes in the 
thermodynamical limit.  
In the annealed calculation, 
the average with respect to $\mbox{\boldmath $\xi$}$ 
is a simple gaussian integral and 
we obtain  the entropy per system size as 
\begin{equation}
{\cal S}^{\rm anneal}(\tilde{\cal M})=
{\alpha}\,{\log}
\left[
1-H(k)
\right]
+{\log}
\left(
2{\cosh}(\tilde{\cal M})
\right).
\end{equation}
The saddle point equation with respect to 
$\tilde{\cal M}$, that is, ${\partial {\cal S}}/{\partial \tilde{\cal M}}=0$ 
leads to $\tilde{\cal M}=0$, and finally we obtain 
\begin{equation}
{\cal S}^{\rm anneal}={\alpha}\,{\log}
\left[
1-2H(k)
\right]
+{\log}2.
\label{ann-s}
\end{equation}
As it is well known, due to the concavity of the 
logarithm, the annealed free energy is a lower bound 
for the correct free energy. 
From this argument and the fact that 
the quenched entropy cannot be negative, 
we obtain a lower bound on the zero temperature 
entropy ${\cal S}^{\rm anneal}$. 
From (\ref{ann-s}), we get ${\alpha}_{\rm c}^{\rm anneal}(k)$, which reads  
\begin{equation}
{\alpha}_{\rm c}(k)\,{\leq}\,-\frac{{\log}2}{{\log}
[1-2H(k)]}={\alpha}_{\rm c}^{\rm anneal}. 
\end{equation}
According to this rigorous condition, 
RS result ${\alpha}_{\rm c}^{(0)}$ 
turns out to be  larger than ${\alpha}_{\rm c}^{\rm anneal}$.
Therefore, we conclude that RS solution is not enough 
to estimate ${\alpha}_{\rm c}$ correctly and more steps of RSB are needed.
We plotted ${\alpha}_{\rm c}^{\rm anneal}$ in FIG. 1.
For large $k$ we get 
${\alpha}_{\rm c}^{\rm anneal}(k)\,{\simeq}\, 
\left(
\sqrt{2/{\pi}}
{\log}2
\right)
k{\rm e}^{k^{2}/{2}}$, and for small $k$
$
{\alpha}_{\rm c}^{\rm anneal}(k)\,{\simeq}\,
-{\log}2/{\log}\left(\sqrt{2/\pi}k
\right)$. 

 In order to estimate the critical number of the noise source, 
${\alpha}_{\rm c}$, 
we may use the another criterion, that is, 
${\alpha}_{\rm s}$ at which 
the zero temperature entropy of the RS saddle point becomes 
negative. 
By numerical calculations, we checked that 
the entropy $S$ of the RS saddle point 
becomes negative at $q=\tilde{q}=0$ for all $k$. 
This  leads to 
\begin{eqnarray}
{\alpha}_{\rm s}={\alpha}_{\rm c}^{\rm anneal}
\end{eqnarray}
and this RS solution is locally stable. 
Therefore, we conclude that 
${\alpha}_{\rm s}$ gives one of the good approximations for 
the critical number of the noise source. 
   
\section{Parisi solution} 

As in \cite{Mourik} the finite step RSB solution is found to be unstable 
against traverse fluctuations, signaling the necessity of a full RSB 
solution. Analytical results can be obtained only  close to  the AT line, 
where the order parameters are small ($q_{\alpha\beta}\sim 
(\alpha-\alpha_{AT})/\alpha_{AT})$, and then the free energy, or 
equivalently the saddle point equations,  
can be expanded in terms of $q_{\alpha\beta}$ and $\tilde{q}_{\alpha\beta}$. 
The expressions involved in here are quite similar to the SK model, where 
for small value of the order parameter, the free energy has, besides 
a quadratic term, a cubic term $\mbox{Tr}q^3$ and a quartic term 
$\sum_{\alpha\beta} q_{\alpha\beta}^4$, whose presence allows a full-fledged 
replica symmetry breaking solution\cite{Parisi2}. 
It is straightforward to see that the saddle point equations, obtained by 
varying (\ref{replica_free_energy}) with respect to 
$q_{\alpha\beta}$ and $\tilde{q}_{\alpha\beta}$, 
coincide with the equation  for the continuum $\mbox{\boldmath $J$}$ case, thus giving 
the same Parisi solution, {\it i.e.} linear up to a breakpoint and then constant$q=q_{EA}$, for any value of $k$ as long as  $k^2 q_{EA}\ll 1$.

Finally, it is worth noticing that the saddle point equation  
with respect to $\tilde{q}_{\alpha\beta}$,
involving the distribution of the couplings $\mbox{\boldmath $J$}$, gives, 
in the proximity of the AT line, $q_{\alpha\beta}\sim\tilde{q}_{\alpha\beta}$ 
for any kind of symmetric distribution of the coupling vectors $\mbox{\boldmath $J$}$ . 
Therefore we conclude that the Parisi solution just above the AT line is an universal function, whereas sizable differences may arise far away from the AT line.

\section{Acknowledgment} 
The authors acknowledge stimulating discussion with H. Nishimori, J. van Mourik
and K. Y. M. Wong. 

D.M.C.'s research was supported by the JSPS under Grant No. P96215.


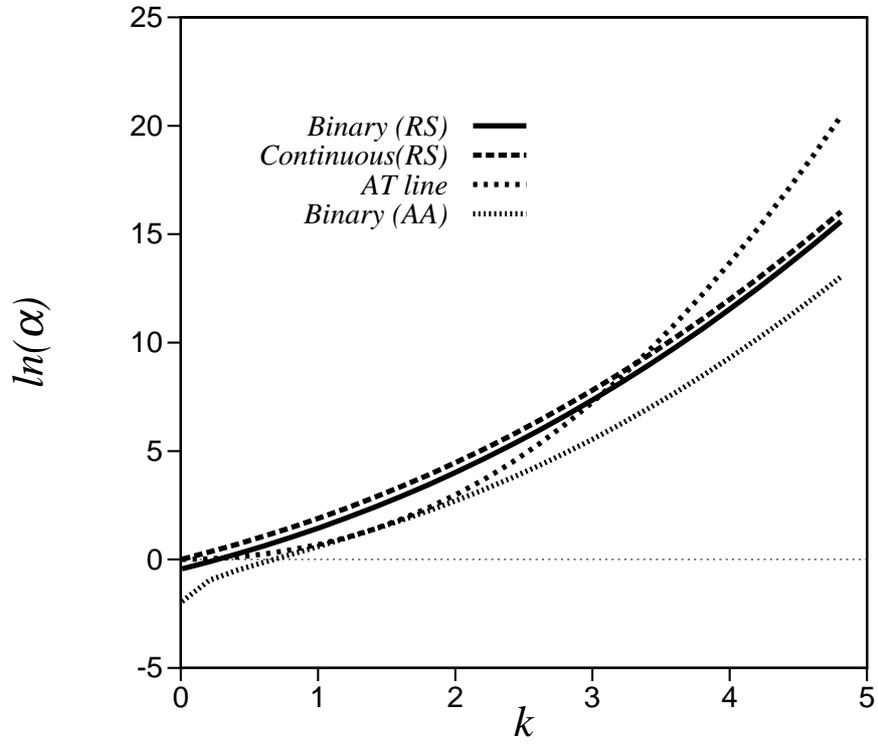
\begin{figure}
\caption{
RS solution, annealed approximation (AA), AT line and 
RS solution by Mourik {\it et al}.
} 
\end{figure}


\begin{references}
\bibitem{Mourik}
J. van Mourik, K. Y. M.Wong and D. Bolle, submitted to {\it Phys. Rev. Lett} 
(cond-mat/9804115).
\bibitem{Gardner}
E. Gardner and B. Derrida, {\it J. Phys. A: Math. Gen.} {\bf 21} (1988) 271.
\bibitem{Parisi} 
M. M\'ezard, G. Parisi and M. A. Virasoro, {Spin Glass Theory and Beyond} 
(World Scientific, Singapore, 1987)
%
%
\bibitem{Krauth}
W.  Krauth and M.  M$\acute{\rm e}$zard, 
{\it J. Phys. France} {\bf 50} (1989) 3057( and references therein).

\bibitem{Parisi2} 
G. Parisi, {\it  J. Phys. A: Math. Gen.} {\bf 13} (1980) L115, 1101, 1887.

\end{references}
\end{document}